# Polarization rotation enhancement by high-order waveguide mode coupling in magnetophotonic crystals

Miguel Levy
Physics Department, Michigan Technological University, Houghton, MI 49931

Rong Li
Materials Science and Engineering Department, Michigan Technological University, Houghton, MI 49931

ABSTRACT

Inter-modal coupling in photonic bandgap optical channels in magnetic films is used to leverage the transverse-electric (TE) to transverse-magnetic (TM) mode conversion due to the Faraday Effect. The underlying mechanism is traced to the dependence of the grating coupling-constant on the modal refractive index and the mode profile of the propagating beam, a feature that arises in waveguide propagation and is not present in normal-incidence stack magnetophotonic crystals. Large changes in polarization near the band edges are observed in first and second orders as a result of the photonic bandgap structure. Extreme sensitivity to linear birefringence exists in second order.

Magneto-optic photonic bandgap structures have been shown to enhance the Faraday rotational response of a material system through photon trapping.[1-4]   Here we report on leveraging the resonant quality of magnetophotonic crystals through high-order-mode coupling in photonic crystal waveguides to further enhance the polarization rotation.   In particular, we show that fundamental-mode propagation in the forward direction can be combined with backscattering into higher-order waveguide modes through the photonic bandgap-structure to achieve a stronger polarization rotational effect while preserving the fundamental-mode forward output.

One dimensional magneto-optic photonic crystals composed of layered stacks operating at normal incidence to the film have been reported by a number of authors.[1,2] Recently, we reported the fabrication of magnetophotonic crystals on ridge waveguides for in-plane propagation, and discussed the effects of linear birefringence on the optical response.[3-5] The present work shows that inter-modal coupling in photonic bandgap optical channels in magnetic films is an important mechanism that can be used to leverage the transverse-electric (TE) to transverse-magnetic (TM) mode conversion due to the Faraday Effect.  We trace the underlying mechanism to the dependence of the grating coupling constant on the modal refractive index and the mode profile of the propagating beam, a feature that arises in waveguide propagation and is not present in normal-incidence stack magnetophotonic crystals.   Large changes in polarization near the band edges are observed in first and second orders, possibly due to strong modifications in normal mode group velocities as a result of the photonic bandgap structure.  Extreme sensitivity to linear birefringence exists in the polarization response in second order.

Six-μm and 7-μm wide ridges fabricated on a 2.86μm-thick liquid-phase-epitaxy (LPE)-grown $(Bi,Lu)_{(2.8 \pm 0.1)}Fe_{(4.6 \pm 0.2)}O_{(12.4 \pm 0.3)}$ film by standard photolithography and ion beam etching serve as the optical channels for this study.  The waveguides consist of a single layer film on (111)-oriented $Gd_3Ga_5O_{12}$ (GGG) with both facets polished. Single-defect (phase-shift step) one-dimensional photonic crystal gratings are fabricated onto the ridge waveguides by focused ion beam (FIB) milling controlled by a Nabity nanometer pattern generation system (NPGS).  The phase-shift step is 3.5 grating-periods long and is placed in the middle of the structure.  Gratings are milled with a FIB aperture of 100μm, beam energy of 30kV and a beam current in the range of 600-800 pA.  Figure 1 shows a partial top view of a photonic crystal on a waveguide, and the inset shows the cross section of the grooves, typically 600nm-deep.  The photonic crystal is 200μm long, and is positioned 20 μm away from the input facet to ensure minimal distortion of the input polarization before the optical beam reaches the photonic crystal.  Here we report on measurements performed on two samples corresponding to waveguides of different widths and lengths: Sample 1 with 7-μm-wide and 1.2-mm-long ridges and Sample 2 with 6-μm-wide and 1.8-mm-long ridges.  Both samples are treated in a post-fabrication cleaning-etch of ceric-ammonium nitrate and perchloric acid solution for 12h to remove re-deposited debris from the FIB process. Other samples tested from the same batch exhibit comparable optical performance to the results reported herein.

Linearly polarized light from a tunable 1480-1580nm wavelength laser source is coupled into the waveguide by end-fire fiber coupling through a lensed fiber. Detection takes place by imaging the output signal into a photodetector through a 10× microscope objective. A saturating 50 Oe magnetic field is applied collinear to the waveguide axis and measurements are taken for parallel and anti-parallel directions of the field to the propagation direction. Elliptically polarized light emerges from the output end due to the birefringent character of the optical channel.[3-5] Herewith the polarization direction is defined as the direction of the semi-major axis of the ellipse, and the polarization rotation angle as half the difference between the output polarization directions corresponding to both magnetic fields. Insertion losses are estimated at 6dB and 7B for samples 1 and 2, respectively. These losses encompass absorption and scattering but not coupling losses.

Figure 2 shows the net transmittance and polarization rotation spectra for Sample 1 for transverse-electric (TE) polarized incident light. The spectral response of the output polarization is plotted for the two magnetic field directions. Notice the reversal in polarization direction for opposite fields. Corresponding ellipticities, defined as the ratio of semi-minor to semi-major axes of the polarization ellipse, are displayed in the inset to this figure. Several stop bands are visible. This stopband multiplicity is not due to high-order Bragg scattering but to high-order waveguide mode coupling through backscattering. Second-order Bragg diffraction occurs well outside the 1480-1580nm wavelength range under study. Beam-propagation simulations and experimental output mode profile analysis show that most of the propagating power (96%) is coupled into the fundamental TE mode in the forward direction with the fiber pointed to the center of the input facet. The spectral dependence of the observed stopbands is not very sensitive to small lateral or vertical displacements of the input fiber across the facet.

The Bragg condition for optical waveguide modes is given by $\lambda = \Lambda(n_f + n_b)$. Here $\lambda$ is the optical wavelength in vacuum, $\Lambda$ the grating period, and $n_f$ and $n_b$ are the modal effective indices of the forward and backward propagating beams, respectively. For a surface relief structure such as in Fig. 1 the mode index is an average quantity that depends on film thickness as well as groove depth and shape. In this work the grooves fabricated by FIB are shaped as shown in the inset to Fig.1.

Prism-coupling measurements on the slab before surface patterning reveal modal indices 2.318, 2.280 and 2.214 for the fundamental, first-, and second-order TE modes, at 1543nm wavelength, respectively. The Faraday rotation per unit length in the material is 137°/mm. Linear-birefringence ($n_{TE} - n_{TM}$) is measured at 0.0006, 0.0046, and 0.0106 for the first three modes. The expression $\lambda = \Lambda(n_f + n_b)$ yields mode indices $n_f = n_b = 2.311$ (Sample 1) and 2.320 (Sample 2) for fundamental mode backscattering from the measured center-wavelength of the deepest infrared stopband in the transmittance spectrum. The difference in effective indices is accounted for by film thickness differences in the presence of the grating grooves and small material index modifications due to the focused-ion-beam processing. The grating period used in these calculations is $\Lambda = 338$ nm as determined by scanning electron microscopy. From the measured spectral values of the first- and second-order backscattered modes center-wavelengths for both samples one can similarly compute the corresponding indices to be $2.262 \pm 0.013$ and $2.182 \pm 0.021$, respectively. Estimates computed from prism coupling data and the standard dispersion relation for TE modes for typical effective-thickness after FIB processing yield $2.263 \pm 0.001$ and $2.178 \pm 0.001$, for the first- and second-order backscattered modes, evincing very good agreement with the backscattering model presented here.

The polarization rotation increases with backscattered mode order, as shown in Fig. 2, displaying very large values of several tens of degrees in high-order. Coupled-mode theory demands an increase in grating-coupling strength with modal order and this stronger coupling is responsible for the large rotations.[6] The effect derives from the coupling of TE and TM modes of different order through the Bragg grating. It should be stressed that in the case of collinear coupling, as we have here, TE to TM mode coupling is not allowed in non-magnetic waveguides (or more generally, if the dielectric tensor has no off-diagonal components),[6] so that the resulting polarization rotation is due to the action of the Faraday Effect modulated by the existing birefringence. In other words, it is the off-diagonal components of the dielectric tensor in the magnetic-garnet that trigger the polarization rotation.

The grating coupling constant between TE modes in a rectangular-profile surface relief grating is given by:[6]

$$\kappa_{TE_n TE_m} = \frac{4\pi}{\lambda} \frac{\sin(a\pi)}{\pi} \frac{h}{\sqrt{T_{effn} T_{effm}}} \frac{\sqrt{(n_f^2 - N_n^2)(n_f^2 - N_m^2)}}{\sqrt{N_n N_m}}$$

where $n$ and $m$ denote the order of the modes, $a$ is the un-etched length fraction of the grating period, $h$ is one-half the groove depth, $n_f$ is the film index, $N_m$ is the effective index corresponding to mode $m$, and $T_{effn}$ is the effective waveguide thickness for guided mode $n$. The latter is equal to the film thickness plus the modal penetration depths into the substrate and cover and depends on the cross-sectional shape of the modes.[6] From this expression we compute coupling constant values 0.006 μm$^{-1}$, 0.012 μm$^{-1}$ and 0.018 μm$^{-1}$ between the fundamental forward mode and the fundamental, first and second-order backscattered modes, respectively. For sinusoidal relief profiles these values are about 20% lower. Notice the increase in coupling strength with modal order, a trait that also holds true for TM to TM mode coupling.[6]

Figure 3 plots the measured polarization rotation for both samples in this study for all three backscattered mode orders considered here. Theoretical predictions are plotted alongside. The theoretical model folds in the effects of grating-coupling strength, Faraday rotation and birefringence into a transfer-matrix calculation.[7] The magnetophotonic crystal waveguide is modeled as a 200μm-long stack of alternating quarter-wave plates with a seven-quarter-wave phase shift in the middle. The index contrast in the stack is adjusted to reproduce the grating coupling-constant calculated from coupled-mode theory for the three different coupling processes. Use is made of the fact that a (non-absorbing) Bragg reflector of length L with coupling constant $\kappa$ yields a transmittance T = 1- tanh$^2$($\kappa$L). Phase changes are preserved in the model by taking the average refractive index in the stack equal to $(n_f + n_b)/2$, with $n_b$ equal to the effective index of the fundamental, first and second order modes, thus satisfying the Bragg condition in each case. The specific Faraday rotation of the garnet and the birefringence enter into the model through the off-diagonal and diagonal components transfer-matrix dielectric tensor, respectively. Good agreement is found with the experimentally measured polarization rotation. The model reproduces the polarization rotation increase with modal order to within the uncertainty of the problem. The uncertainty in the modeled-rotations reflects the degree of uncertainty in the determination of the coupling constant. This is particularly noticeable in the second-order mode where small changes in coupling constant have a significant effect due to the large birefringence of the mode.

M. Levy gratefully acknowledges support from the National Science Foundation under grant ECS 0520814. The authors thank M. J. Steel and A. A. Jalali for a careful reading and comments on the manuscript and H. Dötsch for the LPE samples.

REFERENCES

1. M. Inoue, K. Arai, T. Fujii, and M. Abe, J. Appl. Phys. **85**, 5768 (1999).

2. S. Kahl and A.M. Grishin, Appl. Phys. Lett. **84**, 1438 (2004).

3. R. Li and M. Levy, Appl. Phys. Lett. **86**, 251102 (2005).

4. R. Li and M. Levy, Appl. Phys. Lett. **87**, 269901 (2005).

5. M. Levy, J. Appl. Phys, **99** 073104 (2006).

6. H. Nishihara, M. Haruna, T. Suhara, *Optical Integrated Circuits* (New York, McGraw-Hill, 1985).

7. M. J. Steel, M. Levy, and R. M. Osgood, Jr., J. Lightwave Technol. **18**, 1297 (2000).
9

FIGURE CAPTIONS

Figure 1. Scanning-electron-micrograph of a section of a photonic crystal on magnetic garnet film fabricated for the experiments reported here. Inset: Cross-sectional view of a grating fabricated on LPE-grown $(Bi,Lu)_{(2.8 \pm 0.1)}Fe_{(4.6 \pm 0.2)}O_{(12.4 \pm 0.3)}$ under the same conditions used for the magnetophotonic crystals.

Figure 2. Net transmittance (all polarizations), polarization rotation (solid black line) and direction of the semi-major axis of the output polarization for applied magnetic fields parallel and anti-parallel to the propagation direction (triangles and diamonds). Notice the reversal in polarization direction with field. The polarization rotation is nearly coincident with the diamonds, indicating nearly symmetric polarization reversal. The inset displays the ellipticity defined as the ratio of semi-minor to semi-major axis of the polarization ellipse. The transmittance has been re-plotted for reference.

Figure 3. Maximal polarization rotations (at the wavelengths of largest rotation) for transmitted light for backscattering to fundamental, first and second-order modes. The plot displays the experimental results for Samples 1 and 2 (solid circles) as well as calculated values from transfer matrix simulations for rectangular- (light diamonds) and sinusoidal-grating (black diamonds) coupling constants. The diamonds have been displaced along the horizontal axis for clarity. Notice the considerable increase in rotation with modal order and large variability in second order. The latter is due to sensitivity of the polarization rotation to strong linear birefringence. The error bars denote expected variation due to a groove depth uncertainty of ±50 nm. The inset shows detail for the fundamental mode response.

FIGURES

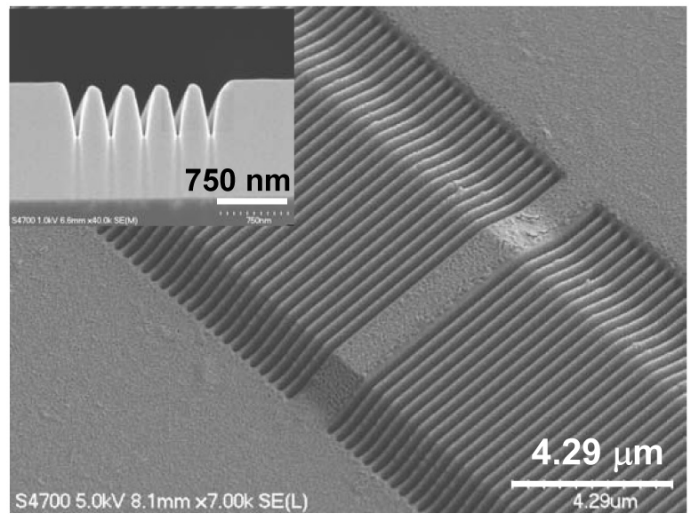

Fig. 1

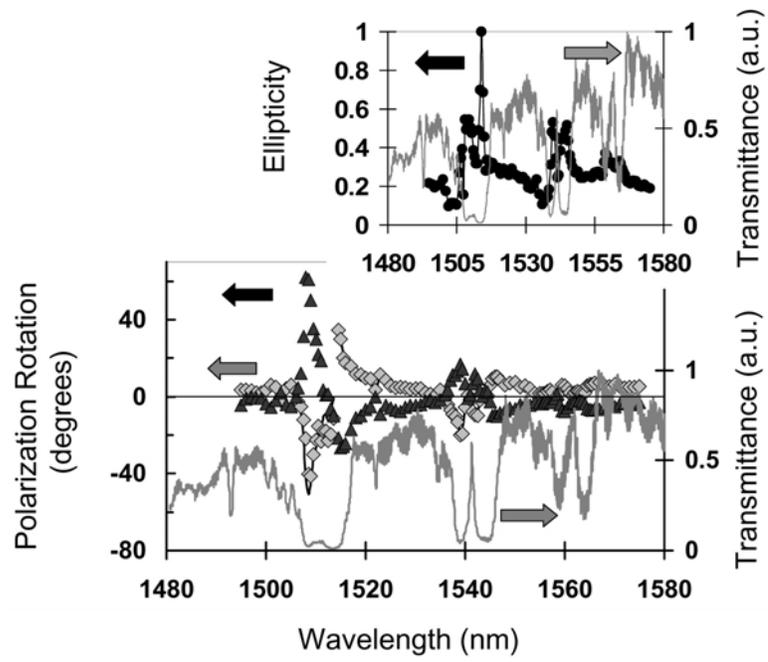

Fig. 2

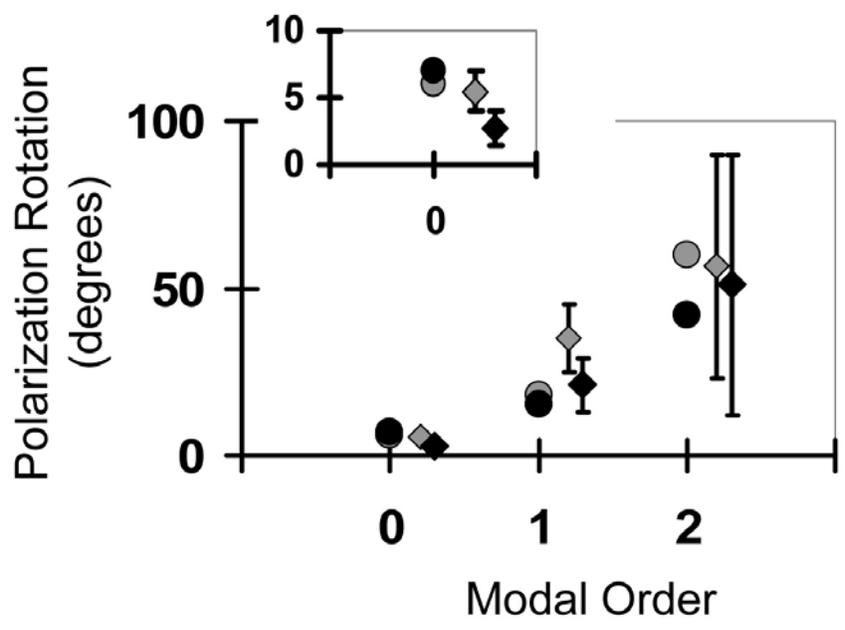

Fig. 3